\documentclass{vgtc}                          

\ifpdf
  \pdfoutput=1\relax                   
  \usepackage{graphicx}                
  \DeclareGraphicsExtensions{.pdf,.png,.jpg,.jpeg} 
\else
  \usepackage{graphicx}                
  \DeclareGraphicsExtensions{.eps}     
\fi%

\graphicspath{{figures/}{pictures/}{images/}{./}} 

\usepackage{microtype}                 
\PassOptionsToPackage{warn}{textcomp}  
\usepackage{textcomp}                  
\usepackage{mathptmx}                  
\usepackage{times}                     
\usepackage{cite}                      
\usepackage{tabu}                      
\usepackage{enumitem}                  
\usepackage{booktabs}                  
\usepackage{booktabs}
\usepackage[table,xcdraw]{xcolor}
\usepackage[normalem]{ulem}
\useunder{\uline}{\ul}{}
\usepackage{longtable}

\onlineid{1064}

\vgtccategory{Research}

\vgtcinsertpkg

\definecolor{mred}{rgb}{.80,.12,.30}
\definecolor{MRED}{rgb}{.80,.12,.30}
\definecolor{grey}{rgb}{0.5,0.5,0.5}
\definecolor{lgrey}{rgb}{0.7,0.7,0.7}
\definecolor{purple}{rgb}{.75,0,.85}
\definecolor{pistachio}{rgb}{0.58, 0.77, 0.45}
\definecolor{myorange}{rgb}{0.94, 0.36, 0.13}

\newif\ifnotes
\notestrue

\let\origcite\cite
\renewcommand{\cite}[1]{\ifnotes\mbox{\origcite{#1}}\else \origcite{#1}\fi}

\title{What is Visualization for Communication? Analyzing Four Years of VisComm Papers}

\author{Vedanshi Chetan Shah\thanks{e-mail: shah.ve@northeastern.edu}\\ %
        \scriptsize Northeastern Univeristy %
\and Ab Mosca \thanks{e-mail: amosca@westfield.ma.edu}\\ %
     \scriptsize Northeastern University} %

\abstract{With the introduction of the Visualization for Communication workshop (VisComm) at IEEE VIS and in light of the COVID-19 pandemic, there has been renewed interest in studying visualization as a medium of communication. However the characteristics and definition of this line of study tend to vary from paper to paper and person to person. In this work, we examine the 37 papers accepted to VisComm from 2018 through 2022. Using grounded theory we identify nuances in how VisComm defines visualization, common themes in the work in this area, and a noticeable gap in DEI practices.     
} 

\CCScatlist{
  {visualizations for communication}
}

\begin{document}


\firstsection{Introduction}

\maketitle


Visualization as a field started with a primary focus on communicating. For example, in 1854 John Snow communicated the source of a cholera outbreak by creating a dot plot on a map~\cite{Snow:1855:MOCC}. And in 1858 Florence Nightingale famously used a rose diagram to communicate the importance of sanitary conditions in the treatment of battle wounds~\cite{Nightingale:1858:DCM}. Although research has shifted away from studying visualization purely as a means of communication, in practice this often continues to be how visualizations are used~\cite{viscomm:2018:VFC}. Moreover, in light of the COVID-19 pandemic, we are seeing a shift towards research that seeks to understand how visualization exists within the greater context of public communication. 

Although visualizations for communication are some of our earliest examples of visualization, there are few theories or guidelines specific to visualizations for communication. As a first step towards addressing this gap, we surveyed all VisComm workshop proceedings from 2018 to 2022. We looked at how these papers define visualization, and found an emphasis on centering the audience, communication, and accessibility. We sorted papers into four categories (new visualizations, theory, guideline, position) and looked for unifying themes within each. We found the importance of iterative design and considering the end user to be common themes across categories, and found themes specific to each category. Notably, despite many papers defining visualization for communication with an emphasis on inclusivity, only three of the 37 papers surveyed offered concrete remarks on Diversity, Equity, and Inclusion (DEI).





\section{Methodology}
We examined the 37 papers published in the VisComm workshop from 2018 to 2022. Our analysis process involved two rounds. In the first round, one of the authors categorized and summarized each paper, establishing initial paper-type categories. Following this, the second author reviewed categories and through collaborative discussions, any disagreements in categorization were resolved. Ultimately, papers were categorized into four types: \textbf{Guideline} papers that present design guidelines\cite{amabili_sultanum_2021, riahi_watson_2021, zhang_sun_padilla_barua_bertini_parker_2020, lévesque_godbout_robert-angers_hurtut_2020, schwabish_feng_2020, jänicke_2019, dasgupta_2018, brath_matusiak_2018}, \textbf{New Visualization} papers that introduce novel visualizations\cite{wu_ku_cheng_shu_puri_wang_qu_2018, janicke_2018, kumar_burch_kirbanismailova_kloos_mueller_2018, wakita_arimoto_2019, flack_ponto_tangen_schloss_2019,stoiber_grassinger_pohl_stitz_streit_aigner_2019, vardhan_2020, yang_ang_wang_2020, coelho_he_baduk_mueller_2020, benda_zikmund-fisher_sharma_ancker_2020, parker_2021, asgari_hurtut_2021}, \textbf{Position} papers that express subjective viewpoints or opinions\cite{liu_wall_patel_park_2020, livingston_brock_2020, ma_millet_2020, wu_tanis_szafir_2019, berube_2019, Schwabish_2018, Parsons_2018}, and \textbf{Theory} papers that propose theoretical frameworks\cite{ma_2021, mangal_turchioe_park_hai_myers_dugdale_goyal_liu_creber_2021, mosca_ottley_chang_2020, matzen_divis_haass_cronin_2020, millet_cairo_majumdar_diaz_evans_broad_2020, arcia_2019, corbeil_daudens_hurtut_2019, hung_parsons_2018, kumar_burch_brand_castelijns_ritchi_rooks_smeth_timmermans_mueller_2018, filipov_ceneda_koller_arleo_miksch_2018}.

The categorization process was followed by a thematic analysis following grounded theory principles\cite{grounded_theory}. Authors identified recurring themes of interest within and between sets of papers. Through several rounds themes were condensed and finalized. In particular, we focused on (a) the way in which papers that defined visualization did so, and (b) commonalities or differences in the key takeaways of papers within each group.

\section{Definitions of Visualization}

Broadly, visualization is defined as the use of visual representation of data to reinforce human cognition~\cite{cardMackinlay1999}. However, different sub-fields each define visualization slightly differently. For example, visual analytics is the science of integrating automatic analysis and human input through interactive visual interfaces to support analytic processing~\cite{keim2008defining}, while scientific visualization is defined by a specific focus on supporting understanding of scientific phenomenon~\cite{Rosenblum1994ScientificV}. 

While all of the definitions of visualization presented in VisComm proceedings included aspects of the general definition of visualization, we note visualization in this context is defined with a particular emphasis on centering the audience, communication, and understandability. For example, seven papers\cite{dasgupta_2018, yang_ang_wang_2020, wu_ku_cheng_shu_puri_wang_qu_2018, ma_millet_2020, Parsons_2018, wu_tanis_szafir_2019, hung_parsons_2018} specifically mention visualizations for communication being geared towards broad audiences, or needing to engage wide populations. Five papers\cite{dasgupta_2018, yang_ang_wang_2020, wu_ku_cheng_shu_puri_wang_qu_2018, ma_millet_2020, mangal_turchioe_park_hai_myers_dugdale_goyal_liu_creber_2021} discuss visualizations explicitly as a way to communicate. And five others\cite{dasgupta_2018, zhang_sun_padilla_barua_bertini_parker_2020, Parsons_2018, wu_tanis_szafir_2019, matzen_divis_haass_cronin_2020} discuss the importance of visualizations for communication being quickly understandable and accessible to wide audiences.

\section{Common Takeaways from Visualization for Communication Work}

Among all papers, we note recurring themes of iterative design and considering end users. The presence of these themes aligns with the emphasis on centering the audience present in definitions of visualization in this space.      



\subsection{Guidelines} 
Within this category (8 papers) we found \textbf{audience consideration}, a clear \textbf{narrative}, and \textbf{iterative design} to be common themes. Three papers highlight the need to tailor the visualizations to the audience's knowledge, preferences, and needs\cite{amabili_sultanum_2021, lévesque_godbout_robert-angers_hurtut_2020, jänicke_2019}. Several papers (2) recommend structuring visualizations to convey a coherent story and effectively communicate an intended message\cite{lévesque_godbout_robert-angers_hurtut_2020, zhang_sun_padilla_barua_bertini_parker_2020}. And three papers emphasize the need for the design processes to include refinement and revision\cite{amabili_sultanum_2021, jänicke_2019, zhang_sun_padilla_barua_bertini_parker_2020}. Only one paper in this group offered DEI guidelines for visualization\cite{schwabish_feng_2020}. 

\subsection{New Visualization}
Among novel visualizations (12 papers) the importance of \textbf{iterative design} and \textbf{user experience} were discussed often. Similar to guideline papers, many of these papers describe using an iterative design process\cite{wakita_arimoto_2019, asgari_hurtut_2021,  parker_2021, vardhan_2020}. Additionally, multiple papers mention user experience and engagement as an important design consideration. Some emphasize intuitive representations\cite{parker_2021}, while others focus on tactile interactions\cite{flack_ponto_tangen_schloss_2019}, facilitating group discussions\cite{benda_zikmund-fisher_sharma_ancker_2020}, or providing clarity to patients and the public\cite{vardhan_2020}.

\subsection{Position}
Within the position category (7 papers) we found highlighting the significance of \textbf{designer experience and intuition} in visualization design to be a common sentiment. Multiple papers suggest experienced designers can rely on their intuition to make informed design choices\cite{livingston_brock_2020, wu_tanis_szafir_2019, berube_2019, Parsons_2018}. Similar to the findings within the guideline category, we note that only two position papers discuss inclusive visualization design and consider the challenges faced by individuals with intellectual and developmental disabilities (IDDs)~\cite{wu_tanis_szafir_2019, liu_wall_patel_park_2020}.

\subsection{Theory}
Theory papers (10 papers) shared a common theme of \textbf{communication}. Three papers in particular explored topics such as effectively conveying information\cite{filipov_ceneda_koller_arleo_miksch_2018}, telling stories\cite{corbeil_daudens_hurtut_2019}, and communicating complex concepts\cite{ma_2021}. Further, multiple papers in this grouping expanded on \textbf{challenges} that arise in visualization broadly and particularly in visualization for communication. For example, conveying variability or uncertainty\cite{millet_cairo_majumdar_diaz_evans_broad_2020}, dealing with biases\cite{matzen_divis_haass_cronin_2020}, finding effective representations\cite{kumar_burch_brand_castelijns_ritchi_rooks_smeth_timmermans_mueller_2018}, or improving visual aesthetics\cite{arcia_2019}.

\section{Discussion and Future Work}
Most of the themes we identified in each paper group aligned with the nuances we identified in VisComm definitions of visualization. However, there is a distinct lack of discussion on implementing and designing inclusive and accessible visualizations. This stands in direct contrast to the ``audience-centered" and ``understandability'' themes we identified as principle components to visualizations for communication. 
Moving forward, we hope to expand and elaborate on our analysis of VisComm proceedings. In addition, we hope our work can motivate more visualization for communication researchers to pay explicit attention to DEI in their work in order to put the field's theoretical values into actual practice.




\bibliographystyle{abbrv-doi}

\bibliography{main}

\begin{thebibliography}{10}

\bibitem{viscomm:2018:VFC}
Visualization for communication (viscomm).
\newblock Accessed: 2021-03-30.

\bibitem{amabili_sultanum_2021}
L.~Amabili and N.~Sultanum.
\newblock Paper maps: Improving the readability of scientific papers via concept maps, Aug 2021.

\bibitem{arcia_2019}
A.~Arcia.
\newblock Colors and imagery in tailored infographics for communicating health information to patients and research participants, Aug 2019.

\bibitem{asgari_hurtut_2021}
M.~Asgari and T.~Hurtut.
\newblock A design space for visualization onboarding in data-driven stories, Oct 2021.

\bibitem{benda_zikmund-fisher_sharma_ancker_2020}
N.~Benda, B.~Zikmund-Fisher, M.~Manoj~Sharma, and J.~Ancker.
\newblock Making numbers meaningful – improving how we communicate numbers to patients and the public, 2020.

\bibitem{berube_2019}
D.~M. Berube.
\newblock Visual communication and heuristics: Challenges and directions from across the disciplines, Sep 2019.

\bibitem{brath_matusiak_2018}
M.~Brath.
\newblock Automated annotations, 2018.

\bibitem{cardMackinlay1999}
S.~Card, J.~Mackinlay, and B.~Shneiderman.
\newblock {\em Readings in Information Visualization: Using Vision To Think}.
\newblock 01 1999.

\bibitem{coelho_he_baduk_mueller_2020}
D.~Coelho, H.~He, M.~Baduk, and K.~Mueller.
\newblock Eating with a conscience: Toward a visual and contextual nutrition facts label, Aug 2020.

\bibitem{corbeil_daudens_hurtut_2019}
J.-P. Corbeil, F.~Daudens, and T.~Hurtut.
\newblock Data visualization + scrollytelling for election news stories : Challenges and perspectives, Sep 2019.

\bibitem{dasgupta_2018}
Dasgupta.
\newblock Bridging computation and visual communication of change using levels of abstraction, 2018.

\bibitem{filipov_ceneda_koller_arleo_miksch_2018}
V.~A. Filipov, T.~Wien, D.~Ceneda, M.~Koller, A.~Arleo, and S.~Miksch.
\newblock The circle of thrones : Conveying the story of game of thrones using radial infographics.
\newblock 2018.

\bibitem{flack_ponto_tangen_schloss_2019}
S.~Flack, K.~Ponto, T.~Tangen, and K.~B. Schloss.
\newblock Lego as language for visual communication, Aug 2019.

\bibitem{hung_parsons_2018}
P.~Hung.
\newblock Affective engagement for communicative visualization: Quick and easy evaluation using survey instruments, 2018.

\bibitem{janicke_2018}
S.~Janicke.
\newblock Time-based impact mosaics, 2018.

\bibitem{jänicke_2019}
S.~Jänicke.
\newblock A visualization course for journalism students, Aug 2019.

\bibitem{keim2008defining}
D.~Keim, G.~Andrienko, J.-D. Fekete, C.~Görg, J.~Kohlhammer, and G.~Melançon.
\newblock Visual analytics: Definition, process, and challenges.
\newblock 03 2008.

\bibitem{kumar_burch_brand_castelijns_ritchi_rooks_smeth_timmermans_mueller_2018}
A.~Kumar, M.~Burch, L.~A. Castelijns, F.~Ritchi, N.~Timmermans, and K.~Mueller.
\newblock Eye tracking for exploring visual communication differences.
\newblock 2018.

\bibitem{kumar_burch_kirbanismailova_kloos_mueller_2018}
A.~Kumar, M.~Burch, D.~Kurbanismailova, U.~Kloos, and K.~Mueller.
\newblock Petalvis-floral visualization for communicating set operations.
\newblock 2018.

\bibitem{liu_wall_patel_park_2020}
S.~Liu, E.~Wall, S.~A. Patel, and Y.~Park.
\newblock Covid-19 health equity dashboard - addressing vulnerable populations, Aug 2020.

\bibitem{livingston_brock_2020}
M.~A. Livingston and D.~Brock.
\newblock Position: Visual sentences: Definitions and applications, Aug 2020.

\bibitem{lévesque_godbout_robert-angers_hurtut_2020}
F.~Lévesque, L.~Godbout, M.~Robert-Angers, and T.~Hurtut.
\newblock How do taxes, benefits and public spending evolve for a taxpayer during their lifetime ?, Aug 2020.

\bibitem{ma_2021}
Q.~Ma.
\newblock Unfolding the decision-making dynamics of news visualization production in china, Aug 2021.

\bibitem{ma_millet_2020}
Q.~Ma and B.~Millet.
\newblock Analyzing dorian twitter data to understand how hurricane risk communication changes as threats unfold, Aug 2020.

\bibitem{mangal_turchioe_park_hai_myers_dugdale_goyal_liu_creber_2021}
S.~Mangal, M.~R. Turchioe, L.~Park, Y.~Hai, A.~C. Myers, L.~Dugdale, P.~Goyal, L.~G. Liu, and R.~M. Creber.
\newblock Know your audience: Comprehension of health information varies by visual format, Aug 2021.

\bibitem{matzen_divis_haass_cronin_2020}
L.~Matzen, K.~Divis, M.~Haass, and D.~Cronin.
\newblock Variable biases: A study of scientists’ interpretation of plot types commonly used in scientific communication, Aug 2020.

\bibitem{millet_cairo_majumdar_diaz_evans_broad_2020}
B.~Millet, A.~Cairo, S.~Majumdar, C.~Diaz, S.~D. Evans, and K.~Broad.
\newblock Beautiful visualizations slain by ugly facts: Redesigning the national hurricane center’s ‘cone of uncertainty’ map, Aug 2020.

\bibitem{mosca_ottley_chang_2020}
Mosca, A.~Ottley, and R.~Chang.
\newblock Does interaction improve bayesian reasoning with visualization?, Aug 2020.

\bibitem{Nightingale:1858:DCM}
F.~Nightingale.
\newblock Diagram of causes of mortality in the army of the east, 1858.

\bibitem{parker_2021}
M.~J. Parker.
\newblock Visualizing student behavior and performance in an online course, Aug 2021.

\bibitem{Parsons_2018}
Parsons.
\newblock Conceptual metaphor theory as a foundation for communicative visualization design, 2018.

\bibitem{riahi_watson_2021}
M.~Riahi and B.~Watson.
\newblock Aesthetics for communicative visualization: a brief review, Aug 2021.

\bibitem{Rosenblum1994ScientificV}
L.~J. Rosenblum.
\newblock Scientific visualization : advances and challenges.
\newblock 1994.

\bibitem{Schwabish_2018}
J.~Schwabish.
\newblock Categorizing and ranking graphs in the american economic review over the last century, Aug 2018.

\bibitem{schwabish_feng_2020}
J.~Schwabish and A.~Feng.
\newblock Applying racial equity awareness in data visualization, Aug 2020.

\bibitem{Snow:1855:MOCC}
J.~Snow.
\newblock On the mode of communication of cholera.
\newblock 1855.

\bibitem{stoiber_grassinger_pohl_stitz_streit_aigner_2019}
C.~Stoiber, F.~Grassinger, M.~Pohl, H.~Stitz, M.~Streit, and W.~Aigner.
\newblock Visualization onboarding: Learning how to read and use visualizations, Aug 2019.

\bibitem{grounded_theory}
Y.~C. Tie, M.~Birks, and K.~Francis.
\newblock Grounded theory research: A design framework for novice researchers.
\newblock {\em SAGE Open Medicine}, 7:2050312118822927, 2019.
\newblock PMID: 30637106.

\bibitem{vardhan_2020}
P.~Vardhan.
\newblock Tile narrative: Scrollytelling with grid maps, Aug 2020.

\bibitem{wakita_arimoto_2019}
K.~Wakita and K.~Arimoto.
\newblock Guiding readers to the focus and context of industrial statistical reports, Sep 2019.

\bibitem{wu_ku_cheng_shu_puri_wang_qu_2018}
A.~Wu, B.~K. Ku, F.~Cheng, X.~Shu, A.~Puri, Y.~Wang, and H.~Qu.
\newblock Pulse: Toward a smart campus by communicating real-time wi-fi access data.
\newblock {\em ArXiv}, abs/1810.00161, 2018.

\bibitem{wu_tanis_szafir_2019}
K.~Wu, S.~Tanis, and D.~Szafir.
\newblock Designing communicative visualization for people with intellectual developmental disabilities, Aug 2019.

\bibitem{yang_ang_wang_2020}
L.~Yang, H.~C. Ang, and S.~Wang.
\newblock Customized visualizations for audience-oriented communication in covid-19 tracking story, Aug 2020.

\bibitem{zhang_sun_padilla_barua_bertini_parker_2020}
Y.~Zhang, Y.~Sun, L.~Padilla, S.~Barua, E.~Bertini, and A.~G. Parker.
\newblock Mapping the landscape of covid-19 crisis visualizations, Aug 2020.

\end{thebibliography}
\end{document}